# LPOR: A Layered Proof of Reserves Framework for Usable and Publicly Auditable Solvency Verification

Donggoo Kim
*College of Computing*
*Georgia Institute of Technology*
Atlanta, USA
dkim3055@gatech.edu

Rajesh Upadhayaya
*Computer Science*
*University of New Mexico*
Albuquerque, USA
rajeshupadhayaya@unm.edu

Milosz Bator
*College of Computing*
*Georgia Institute of Technology*
Atlanta, USA
mbator3@gatech.edu

Tao Le
*College of Computing*
*Georgia Institute of Technology*
Atlanta, USA
tao8@gatech.edu

***Abstract*—Proof of Reserves (PoR) enables centralized crypto exchanges to demonstrate that on-chain reserves are sufficient to cover customer liabilities. However, existing approaches, including Merkle-tree-based proofs and zero-knowledge PoR systems, remain difficult for everyday users to verify in practice, resulting in limited participation and weakened transparency. We introduce *LPOR*, a layered, usability-focused PoR framework that separates lightweight user-side checks from auditor-level cryptographic verification, enabling non-technical users to verify inclusion and publicly recompute total liabilities with minimal friction. By lowering verification barriers, LPOR increases user participation and substantially improves the probability of detecting omitted liabilities. We evaluate its scalability and omission detectability at a multi-million-user scale.**

## I. Introduction

Proof of Reserves (PoR) is intended to let centralized exchanges demonstrate that on-chain reserves are sufficient to cover customer liabilities. While we use the term Proof of Reserves (PoR) following industry convention, LPOR primarily addresses the proof of liabilities component. Custodial platforms remain a primary access point for digital assets, yet past failures such as Mt. Gox and FTX show that users cannot safely treat exchange solvency as implicit [1], [3].

Existing PoR approaches do not fully meet that goal. Address disclosure reveals assets but not liabilities, Merkle-tree PoR supports liability inclusion proofs but is cumbersome for ordinary users, and zero-knowledge PoR improves privacy and expressiveness at the cost of greater implementation and verification complexity. This creates asymmetric verifiability: auditors can assess proofs, while users face high technical barriers.

We propose **LPOR**, a layered PoR framework that separates lightweight checks from auditor-level cryptographic verification. LPOR exposes a public liability ledger built from tokenized balances, allowing users to verify inclusion and recompute total liabilities in a human-readable form, while auditors independently verify that the published ledger is bound to the exchange's Merkle commitment. The system therefore assigns clear roles to the CEX, users, and auditors without requiring every participant to execute cryptographic verification code.

Our contributions are:

- Separation of user and auditor verification roles
- Minimal technical effort for user-side verification
- Publicly verifiable computation of total liabilities
- No disclosure of user-identifiable information

## II. Existing PoR Techniques

### A. Wallet Address Disclosure PoR

Wallet address disclosure provides transparent evidence of on-chain assets, but it does not account for liabilities and therefore cannot establish solvency on its own [8]. It also remains vulnerable to snapshot manipulation such as temporary reserve borrowing, making it an incomplete PoR mechanism [6], [7].

### B. Merkle-Tree-Based PoR

Merkle-tree-based PoR adds user-level liability inclusion proofs while preserving privacy, and it is widely deployed in practice [4], [5], [9]. Its main limitations are poor usability for ordinary users and the inability to publicly recompute total liabilities from the opaque commitment alone [2].

### C. Zero-Knowledge-Proof-Based PoR

Zero-knowledge PoR can prove aggregate solvency while hiding individual balances and avoiding some forms of auditor trust [3], [10]. However, these systems rely on substantially more complex proof infrastructure and remain difficult for non-technical users to inspect or verify directly.

Recent works on Proof of Liabilities (PoL), such as [11], [12], explore privacy-preserving constructions using commitments and zero-knowledge techniques. While these approaches provide strong cryptographic guarantees, they often require specialized verification and introduce significant complexity for end users. LPOR takes a complementary approach by focusing on usability and participation rather than stronger cryptographic expressiveness.

## III. LPOR: Layered Proof of Reserves

### A. Protocol Overview

**System Roles.** LPOR involves three parties: (i) the centralized exchange (CEX), which generates and publishes the proof; (ii) users, who verify inclusion of their balances and check total liabilities; and (iii) auditors, who recompute Merkle commitments to ensure cryptographic binding. LPOR assumes at least one honest auditor and relies on public verifiability rather than trust in any single party.

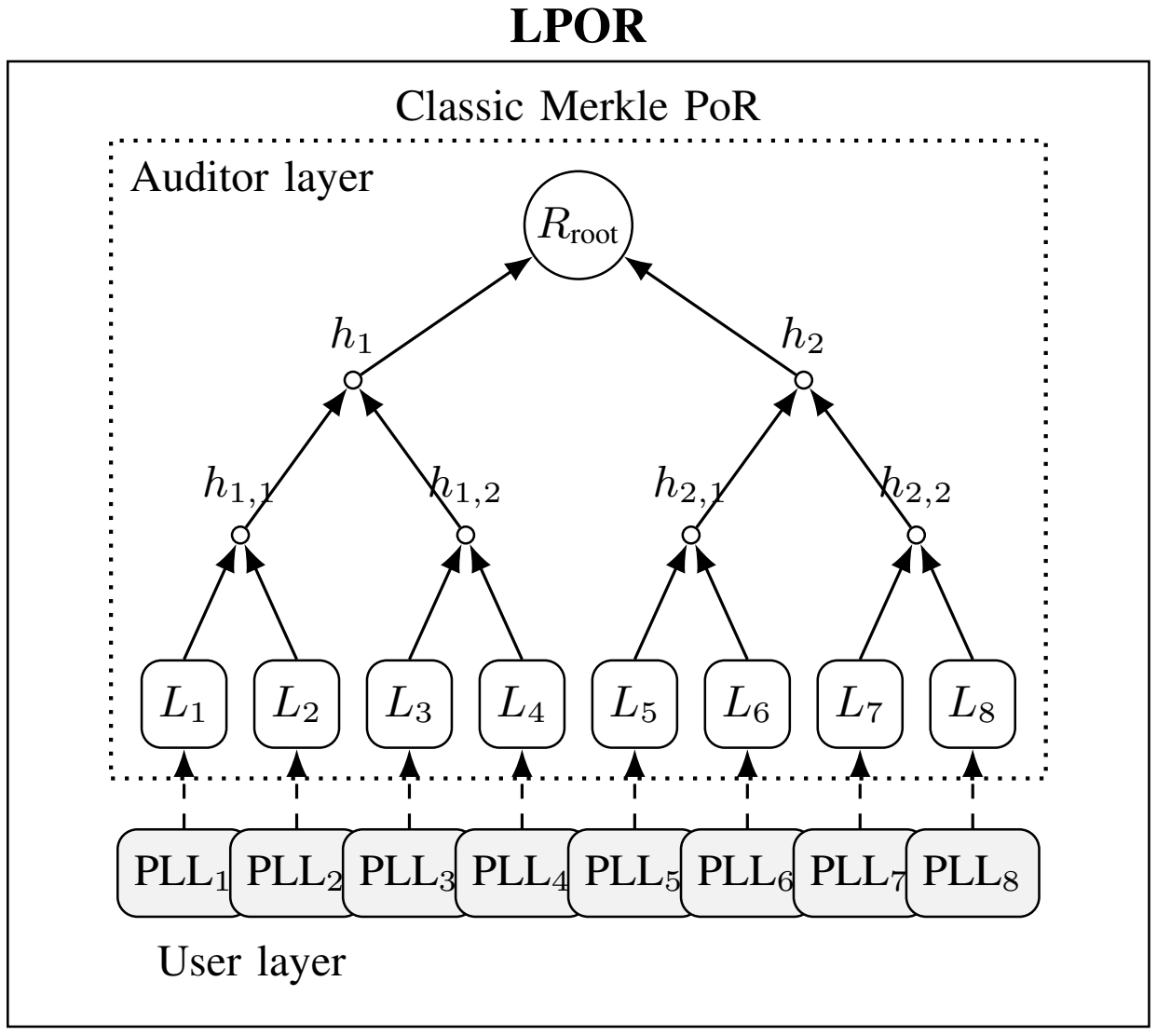

LPOR shifts cryptographic verification from end users to an open auditor layer (i.e., any independent, technically capable party) by decoupling (i) lightweight user checks from (ii) auditor recomputation of binding Merkle commitments, improving usability without weakening the binding property of the Merkle commitment. This separation shifts verification from institution-driven proofs to user-driven validation through simple aggregation.

Specifically, LPOR ensures: (1) individual liability values are publicly accessible in anonymized form, (2) no user-identifiable information is disclosed, (3) the total sum of all user balances is publicly verifiable, and (4) verification tasks are divided into a user-facing layer and a non-interactive auditor layer.

Unlike classical Merkle-tree-based PoR schemes, LPOR does not require a trusted third-party auditor to privately aggregate user balances in order to compute total liabilities.

**Protocol 1: Balance Tokenizer.** The exchange converts each user's balance into standardized denomination tokens (e.g., 0.1 TBTC, 1 TBTC). Each token is assigned a unique identifier (UUID) derived from the user's account information in a privacy-preserving manner.

**Protocol 2: Public Liability Ledger.** The exchange publishes a publicly accessible data table containing the tokenized balances generated by the Balance Tokenizer. This enables users to independently verify the inclusion of their assets and validate the publicly computed total sum of all balances in a human-readable format.

### B. Trust Assumptions and Commitments

For each proof epoch, the exchange publishes (i) the Public Liability Ledger (PLL), (ii) a Merkle root $R$ committing to all PLL records, and (iii) a set of disclosed reserve wallet addresses.

In classical Merkle PoR, the commitment binds to semantically opaque hash values (e.g., a leaf hash such as `688787d8f`), which prevents public recomputation of total liabilities without access to the exchange's internal records.

In contrast, LPOR commits to an explicit, non-negative liability ledger. Each Merkle leaf corresponds to a tokenized balance record of the form `(UUID, token, value)` for example `(688787d8f, TBTC, 0.1)`, where UUID is a deterministic hash over user-specific data, token attributes, and the proof identifier. This preserves cryptographic binding while making total liabilities publicly verifiable via a simple summation over the PLL. Under standard hash collision-resistance assumptions, the published Merkle root $R$ is a binding commitment to the full liability set.

As described in the system model, LPOR relies on open verification, where independent auditors can recompute and validate the published commitments. Verification is deterministic and publicly reproducible under standard cryptographic assumptions.

### C. Balance Tokenizer

The Balance Tokenizer is the foundational component of LPOR that transforms user balances into standardized tokens, enabling public verification of liabilities. Tokenization serves a dual role: (i) enforcing non-negativity by construction, and (ii) enabling public aggregation without revealing exact user balances, which would otherwise be directly exposed in a raw balance list. Tokenization also introduces shared denomination patterns across users, providing limited k-anonymity while preserving public verifiability.

*1) Tokenization Process:* The Balance Tokenizer operates by decomposing each user's BTC balance into standardized denominations using a greedy algorithm. For a given user balance, the system creates tokens in descending order of value: 10 TBTC, 1 TBTC, 0.1 TBTC, and 0.01 TBTC tokens. For example, a user with a balance of 3.12 BTC would receive three 1 TBTC tokens, one 0.1 TBTC token, and two 0.01 TBTC tokens, fully representing their holdings through discrete, standardized units.

$$UUID_i = H(\text{user_id}_i \mid\mid \text{token}_i \mid\mid \text{value}_i \mid\mid \text{proof_id})$$

Each generated token receives a unique identifier (UUID) derived from a cryptographic hash of the user's account information combined with token attributes and a public proof identifier. The `proof_id` is fixed for all tokens within a given proof epoch and changes across snapshots, ensuring that identifiers remain consistent within a proof while differing across distinct proofs.

Users can deterministically derive their identifiers from account information, although in practice the exchange may provide these identifiers to simplify verification.

This hash-based construction enables users to verify token inclusion without revealing their identity to external observers, while preserving privacy and reducing linkage of user balances across different proof epochs.

### *D. Public Liability Ledger*

The output of the Balance Tokenizer is the Public Liability Ledger (PLL), a critical interactive layer that enables publicly verifiable proof of liabilities in a human-readable format.

The Public Liability Ledger (PLL) is a mirrored data table derived from the User Balance Table in the CEX's internal database. This transformation is performed by the LPOR Balance Tokenizer.

TABLE I
PUBLIC LIABILITY LEDGER EXAMPLE

| User (private) | UUID | Token | Value |
|---|---|---|---|
| User_1 | 688787d8f | TBTC | 0.1 |
| User_2 | 5fd924625 | TBTC | 0.1 |
| User_2 | d8a928b20 | TBTC | 1 |
| User_2 | b013907f9 | TBTC | 1 |
| User_2 | 3c95ec524 | TBTC | 1 |
| User_1 | 564e5ede3 | TBTC | 10 |
| User_3 | 9b19be624 | TBTC | 100 |

Users can only verify the association between tokens and their own accounts, as the `User` column is private.

- **Inclusion Proof:** A user confirms that their tokens appear in the PLL.
- **Total Sum Proof:** The total sum of all values in the PLL, with no negative balances, can be computed via a simple summation.

### *E. Public Auditor Verification*

The auditor layer is open: any independent party can recompute and verify the published commitments. Each PLL record is hashed as

$$L_i = H(\text{UUID}_i \,||\, \text{token}_i \,||\, \text{value}_i),$$

and auditors recompute the Merkle tree to verify that the resulting root matches the published commitment.

This process is deterministic and publicly reproducible, enabling independent verification. In practice, such verification is performed by third parties with economic or reputational incentives, including regulators, security firms, and independent researchers.

## IV. THREAT MODEL AND SECURITY GOALS

**Threat Model.** We model the centralized exchange (CEX) as a potentially malicious prover that may misrepresent its solvency. We consider three adversarial behaviors: (i) omission of liabilities, (ii) manipulation via hidden negative balances, and (iii) inconsistent disclosure of liability data across users.

LPOR adopts an open auditing model, where correctness holds as long as at least one independent verifier recomputes and validates the published commitment.

**Security Goals.** LPOR targets the following goals: (G1) *Binding*: a published snapshot cannot be altered without detection under standard hash assumptions; (G2) *Public auditability*: any party can recompute total liabilities and verify consistency with the published commitment; (G3) *Omission detectability*: missing liabilities can be detected with probability increasing in user participation.

**Defense Mechanisms.** *Omission (G3).* LPOR exposes a human-readable Public Liability Ledger (PLL), allowing users to verify inclusion of their balances. Omitted liabilities can therefore be detected when affected users participate.

*Internal offsets (G1).* All liabilities are represented as non-negative standardized tokens, preventing hidden negative balances and enforcing non-negativity by construction.

*Inconsistent disclosure (G1, G2).* LPOR commits to a single Merkle root over the PLL. Any inconsistent views result in mismatched commitments, which are detectable through public recomputation.

**Limitations.** LPOR does not prevent short-term reserve borrowing ("window dressing"), a limitation shared by snapshot-based PoR systems.

## V. EVALUATION

### *A. Qualitative Evaluation*

TABLE II
COMPARISON OF POR METHODS. L. INCLUSION DENOTES USER-VERIFIABLE LIABILITY INCLUSION; L. SUMMATION DENOTES PUBLIC RECOMPUTATION OF TOTAL LIABILITIES.

| Property | Addr. disclosure | Merkle-PoR | zk-PoR | LPOR |
|---|---|---|---|---|
| Assets | O | O | O | O |
| L. Inclusion | X | O | O | O |
| L. Summation | X | X | O | O |
| Verifiability | Public | Partial | Public | Public |
| User Effort | Low | High | High | Low |
| Complexity | Low | Medium | High | Medium |

We compare four PoR approaches along dimensions most relevant to practical transparency and user participation.

**Wallet Address Disclosure** provides public visibility into on-chain assets with minimal user effort, but does not account for liabilities.

**Merkle PoR** introduces cryptographic liability inclusion via Merkle commitments. While users can verify inclusion of their own balances, the opaque commitment prevents public recomputation of total liabilities and imposes high verification effort.

**zk-PoR** enables cryptographic verification of aggregate solvency, including liability summation, without revealing individual balances. However, proof generation and verification rely on specialized cryptographic infrastructure, limiting accessibility for non-technical users.

**LPOR (proposed)** exposes a public, tokenized liability ledger that supports both user-verifiable inclusion and public liability summation through simple checks. It achieves public auditability with low user effort and moderate complexity while maintaining cryptographic binding via Merkle commitments.

### B. Quantitative Evaluation

We evaluate LPOR along deployment-critical scalability axes: (i) exchange-side proof generation cost, (ii) proof size, and (iii) public verification time.

**Experimental setup:** We vary the number of users as the independent variable ($N \in \{10^4, 10^5, 10^6, 10^7\}$). All measurements are *single-threaded* on commodity hardware (macOS, Python 3.11; 16 CPU cores available, 128 GB RAM). Public verification assumes a realistic 100 Mbps download rate. Each proof epoch publishes: (i) the Public Liability Ledger (PLL), (ii) a Merkle root commitment, and (iii) minimal metadata. *All experiments assume a single asset type (BTC).*

TABLE III
QUANTITATIVE SCALABILITY RESULTS (SINGLE-THREADED). "PROOF SIZE" INCLUDES PLL + MERKLE ROOT + METADATA. "PUBLIC VERIFY" INCLUDES DOWNLOAD + RECOMPUTATION (SUM + MERKLE RECOMPUTATION) ASSUMING 100 MBPS.

| Users | Proof Size | Proof Generation | Public Recomputation |
|---|---|---|---|
| $10^4$ | 5.10 MB | 0.69 s | 0.53 s |
| $10^5$ | 51.13 MB | 7.69 s | 5.40 s |
| $10^6$ | 510.97 MB | 81.42 s | 54.96 s |
| $10^7$ | 5.11 GB | 880.89 s | 568.76 s |

Across all scales, LPOR shows predictable linear growth in proof size, proof generation, and public recomputation time. Proof size stays near ≈536 bytes/user, and end-to-end proof generation reaches 10.5 minutes at $10^7$ users. Public recomputation is similarly practical: at $10^7$ users, full verification completes in 9.48 minutes under a 100 Mbps link, with most of the cost due to downloading the PLL rather than cryptographic processing.

**Takeaway:** LPOR preserves public auditability at multi-million-user scale on commodity hardware, with costs dominated by simple linear data processing rather than specialized cryptography.

### C. Omission Detectability Analysis

We analyze how LPOR improves the probability of detecting omitted liabilities by increasing user participation through low-friction verification.

Let $p$ denote the probability that a user performs inclusion verification during a proof epoch, and let $k$ be the number of omitted users. An omission is detected if at least one omitted user verifies inclusion:

$$P_{\text{detect}} = 1 - (1-p)^k.$$

In classical Merkle-tree-based PoR systems, inclusion verification requires executing cryptographic verification scripts, resulting in very low participation in practice. We therefore consider $p \in [0.01\%, 0.1\%]$ for Merkle PoR. LPOR reduces verification to a human-readable check of token inclusion and public summation, enabling substantially higher participation; we evaluate $p \in [0.5\%, 5\%]$.

TABLE IV
OMISSION DETECTABILITY VS. USER PARTICIPATION. ASSUMPTIONS: $N = 10^7$ USERS, OMISSION FRACTION $f = 0.001\%$, OMITTED USERS $k = 100$.

| Participate rate $p$ | Scheme | Detection prob. |
|---|---|---|
| 0.01% | Merkle PoR (conservative) | 0.995% |
| 0.1% | Merkle PoR (optimistic) | 9.52% |
| 0.5% | LPOR (conservative) | 39.35% |
| 1% | LPOR (moderate) | 63.45% |
| 5% | LPOR (optimistic) | 99.41% |

Under these assumptions, LPOR increases omission detectability from below 10% to over 99% without altering the underlying Merkle commitment construction. The improvement arises solely from higher user participation enabled by simplified verification.

## VI. DISCUSSION

LPOR adopts controlled partial disclosure using standardized tokens and UUIDs, reducing direct user identifiability and enforcing non-negativity without zero-knowledge complexity. Compared to zk-PoR, it trades stronger privacy guarantees for human-readable verification.

Its core contribution is the separation of user checks and auditor verification. Users perform inclusion and summation checks, while auditors enforce binding via Merkle recomputation. This removes reliance on opaque auditor reports.

LPOR incurs larger proof sizes due to tokenization overhead, a deliberate trade-off between storage efficiency and verification accessibility, acceptable when transparency is prioritized.

LPOR frames PoR verification as a socio-technical system rather than a purely cryptographic artifact, combining public readability with cryptographic commitments.

## VII. CONCLUSION

This work shows that PoR systems must balance cryptographic soundness with practical verifiability.

LPOR reframes PoR as a layered verification problem, separating low-friction, user-facing checks from auditor-level cryptographic verification. Our evaluation shows that this design enables public recomputation of liabilities at a multi-million-user scale and improves omission detectability without introducing complex zero-knowledge infrastructure.

We view LPOR not as a replacement for existing PoR techniques, but as a complementary design point that highlights the role of usability in financial transparency systems. More broadly, future PoR mechanisms should treat accessibility as a core design parameter alongside traditional cryptographic guarantees.